# Magnetic-Field Control of Tomonaga-Luttinger Liquids in Ta$_2$Pd$_3$Te$_5$ Edge States

Xingchen Guo[1,2,†], Anqi Wang[1,2,†], Xiutong Deng[1,2,†], Yupeng Li[3], Guo'an Li[1,2], Zhiyuan Zhang[1,2], Xiaofan Shi[1,2], Xiao Deng[1,2], Ziwei Dou[1], Guangtong Liu[1], Fanming Qu[1,2], Zhijun Wang[1,2], Tian Qian[1], Youguo Shi[1,2,*], Li Lu[1,2,*], Jie Shen[1,*]

[1]Beijing National Laboratory for Condensed Matter Physics, Institute of Physics, Chinese Academy of Sciences, Beijing 100190, China

[2]School of Physical Sciences, University of Chinese Academy of Sciences, 100049 Beijing, China

[3]Hangzhou Key Laboratory of Quantum Matter, School of Physics, Hangzhou Normal University, Hangzhou 311121, China

†These authors contributed equally to this work

**\*Corresponding author**

e-mail: ygshi@iphy.ac.cn; lilu@iphy.ac.cn; shenjie@iphy.ac.cn

**Abstract**

Ta$_2$Pd$_3$Te$_5$ is a quasi-one-dimensional transition-metal telluride whose heavy atoms endow the material with strong spin-orbit coupling, while the Fermi level inside the bulk gap makes the low-energy electronic structure highly tunable. Theory and early experiments have already identified a wealth of emergent phases in this platform: an excitonic insulator driven by electron-hole binding, a second-order topological insulator protected by crystalline symmetry, a potential topological-protected quantum-spin-Hall edge, and proximity-induced edge supercurrents when coupled to a conventional s-wave superconductor. These properties make it a promising platform for hosting Majorana zero modes and quantum computation, provided that time-reversal symmetry can be broken by a Zeeman gap. In this work, we demonstrate that the one-dimensional edge channels of exfoliated Ta$_2$Pd$_3$Te$_5$ host a robust and tunable Tomonaga-Luttinger liquid by electrostatic gating because it shifts the chemical potential across the bulk gap without changing the gap size. More importantly, the application of a magnetic field introduces a Zeeman gap that systematically increases the TLL power-law exponent $\alpha$. Furthermore, rotating the field reveals a pronounced twofold anisotropy—$\alpha$ is maximal for a field parallel to the edge and minimal for a perpendicular orientation—originating from an orientation-dependent edge *g*-factor that is likely amplified by quantum-confinement-induced orbital-angular-moment quenching. The existence of gate-tunable edge supercurrents together with the field-controlled Zeeman gap provides a direct route to break time-reversal symmetry in a particle-hole-symmetric superconducting gap and thus to engineer a topological superconducting phase, paving the way towards Majorana-based quantum devices.

**Introduction**

One-dimensional (1D) quantum systems do not conform to the Landau Fermi-liquid paradigm. Instead, their low-energy excitations are collective bosonic modes that are captured by Tomonaga-Luttinger liquid (TLL) theory[1-4]. The TLL framework describes a set of universal phenomena—spin-charge separation, power-law correlation functions, and interaction-dependent scaling exponents—that manifest, for example, as a characteristic power-law dependence of the conductance on both bias voltage ($V_{bias}$) and temperature ($T$) in electronic transport experiments. In recent years, the field has broadened dramatically, encompassing a variety of intertwined platforms[5], such as quantum wires and carbon nanotubes[6,7], semiconducting nanowires (InAs, InSb)[8,9], ultracold atomic gases confined to 1D[10], edge states of integer and fractional quantum Hall systems[11,12], and van-der-Waals (vdW) heterostructures[13]. Of particular interest are the helical TLL that arises in spin-orbit-coupled (SOC) nanowires and quantum-spin-Hall edges, as well as the chiral TLL that appears at the boundaries of fractional quantum Hall states and fractional Chern insulators, all are intimately linked to the emergence of Majorana zero modes and to the realization of topological phases[14-18]. In these contexts, the response of a TLL to an external magnetic field—usually described by the Zeeman energy—is crucial because the field induces time-reversal symmetry breaking and spin polarization, a prerequisite for many topological transitions in particular for the helical LL. Consequently, a central question that has emerged is: How does a magnetic field modify spin-charge separation, the linear dispersion, and the gapless nature of a Tomonaga-Luttinger liquid?

$Ta_2Pd_3Te_5$ is a quasi-one-dimensional (Q1D) transition-metal telluride whose structural anisotropy, sizable spin-orbit coupling and proximity to a superconductor make it an intriguing candidate for use in topological phenomena and high-performance superconducting devices. Ongoing experimental and theoretical work aims to reveal materials that can exhibit multiple novel quantum phenomena, including excitonic insulators[19,20], second-order topological insulators[21,22], quantum-spin-Hall states[20] and edge supercurrents[23]. Among the exotic properties, Luttinger liquids have also been observed in edge states and present great extension from micro- to macrosamples[21]. Owing to the fact that the Fermi level is naturally pinned inside the bulk gap, the electron-electron interaction strength can be tuned continuously by electrostatic gating, enabling a controllable crossover from a TLL to a conventional Landau Fermi liquid (LFL) on the same platform[21]. The ease with which $Ta_2Pd_3Te_5$ can be brought into proximity with a conventional superconductor has already yielded robust edge-state supercurrents[23]. This paves the way towards a topological phase transition that requires breaking time-reversal symmetry, typically achieved by Zeeman energy induced by an outer magnetic field[24-26]. Consequently, a systematic investigation of the edge states under applied magnetic fields is both timely and essential for unlocking the full potential of this material.

In this work, we fabricate few-layer $Ta_2Pd_3Te_5$ devices and show that the edge TLL can be tuned not only by electrostatic gating but also by the magnitude and orientation of an external magnetic field. The tuning is quantified through the power-law exponent $α$ extracted from the applied bias voltage and temperature dependence of the differential conductance ($dI/dV$), which directly reflects the strength of electron-electron correlations. Applying a magnetic field monotonically increases $α$, indicating that a Zeeman gap opening in the spin sector is related to an effective $g$-factor value. Remarkably, $α$ exhibits a pronounced twofold anisotropy: it is maximal when the field is parallel to the edge and minimal for a perpendicular field. This behaviour is attributed to an anisotropic $g$-factor that is modified by the Q1D quantum-confined geometry of $Ta_2Pd_3Te_5$. Confinement reduces the orbital angular momentum (OAM) in the transverse direction, enhancing the *g*-factor for fields aligned with the edge direction and thereby producing a larger Zeeman gap. Interestingly, the bulk semiconductor gap remains invariant under both the gating and the magnetic field, confirming that the observed exponent *α* changes are intrinsic to the edge state—a crucial step for engineering the edge state for Majorana cases. Thus, our findings reveal the intertwined roles of electron-electron interactions in 1D physics, SOC, quantum confinement and anisotropic magnetic-field effects, and provide a practical handle for engineering magnetic-field-controlled topological phases and potential Majorana-type devices within a single material system.

**Basic properties of $Ta_2Pd_3Te_5$ devices**

$Ta_2Pd_3Te_5$ is a layered material whose interlayers are coupled through vdW interactions, and each of its layers is constructed of Q1D $Ta_2Te_5$ chains and Pd atoms, as shown in Fig. 1a. A formal study revealed that one type of $Ta_2Pd_3Te_5$ can exhibit three distinct behaviours when *T* changes or $V_{bias}$ is applied, indicating that there are three states in this material[21]. At room temperature or a high $V_{bias}$, the bulk semiconductor state dominates the behaviour of $Ta_2Pd_3Te_5$. As the *T* or $V_{bias}$ decreases, the edge Luttinger liquid state appears gradually, and *dI/dV* decreases linearly according to the log-log plot of the *dI/dV-T* and *dI/dV-$V_{bias}$* curves. If *T* or $V_{bias}$ continues to decrease, the edge gap feature will show up. This typical electronic transport has been studied in multiple devices before[21].

The vdW interlayer interaction makes it easy for $Ta_2Pd_3Te_5$ to be exfoliated into few layers. We utilize this characteristic to prepare few-layer devices, among which device #1 is shown in Fig. 1b and device #2 is shown in Supplementary Fig. 3a. The back gate electrode is 5/15 nm Ti/Au with few-layer hBN acting as the gate dielectric layer, and the source and drain electrodes are 5/50 nm Ti/Au. In device #1, we first examined its three previously reported states and typically the Luttinger liquid properties. We measured $dI/dV$-$V_{bias}$ curves at different $T$ and a -10V back gate voltage ($V_{bg}$=-10 V), which is near the charge-neutral-point (CNP) as shown in Fig. 1c, and then plotted them as scaled conductance $(dI/dV)/T^{0.34}$ versus scaled bias voltage $eV_{bias}/k_BT$, as shown in Fig. 1d. At the relatively high temperature plotted in Fig. 1d, the small edge gap is smeared by the high thermal excitation and does not dominate, so the curve shows a standard power-law relation as a function of $V_{bias}$. Moreover, a suitable coincidence confirms the existence of the Luttinger liquid state, as shown in our previous paper[21]. In contrast, at the much lower temperature shown in Fig. 1e, the $dI/dV$-$V_{bias}$ curves at $V_{bg}$=-10 V presents three typical regimes, which indicate edge gaps, edge Luttinger liquids and bulk semiconductors. This is also confirmed by the $dI/dV$-$T$ curve at $V_{bg}$=-10 V illustrated in Fig. 1f, which shows edge gap behaviour at 4 K.

**Gate-tunable edge Luttinger liquid at zero and finite magnetic field**

Moreover, we found that gating can tune the Luttinger liquid by changing its power-law exponent *α* of the *dI/dV*-*V*<sub>bias</sub> curves and *β* of the *dI/dV*-*T* curves. We measured a series of $dI/dV$-$T$ curves at different $V_{bg}$ (Fig. 2a), and all the curves coincide at high temperature, indicating that gating has little influence on the bulk state because of the large thermal excitation around the Fermi level and the broadening of the Fermi-Dirac distribution. As the temperature decreases, the $dI/dV$-$T$ curves at $V_{bg}$<6 V transit to the edge gap state after going through the edge Luttinger liquid state, whereas those at $V_{bg}\geq$ 6 V still maintain the edge Luttinger liquid state down to the lowest temperature of ~1.6 K. The power-law exponent *β* extracted from the power-law regime varies from 0.14 to 0.34, as shown in Fig. 2b, revealing that the closer to the CNP, the greater its value, as reported previously[21].

We measured a series of $dI/dV$-$V_{bias}$ curves at 1.6 K and different $V_{bg}$ values (Fig. 2c). Similar to the $T$-dependent measurement, the $dI/dV$-$V_{bias}$ curves at $V_{bg}$<6 V will first go through the edge gap state and then activate the edge Luttinger liquid behaviour as $V_{bias}$ increases, whereas curves measured at $V_{bg} \geq 6$ V will enter into the edge Luttinger liquid state directly after a plateau at small $V_{bias}$ because of thermal excitation and AC excitation. This finding indicates that when $V_{bg} \geq 6$ V is applied, the Fermi level will be lifted above the edge gap and is located at the gapless edge conducting state, which is why the TLL appears even at a small $V_{bias}$. By extracting the power-law exponent $\alpha$ from these $V_{bias}$-dependence curves, we found that the closer to the CNP, the higher the $\alpha$, as shown by the blue dots in Fig. 2d. This is the same as the $T$-dependence measurement in Fig. 2b. We noticed that the $V_{bias}$-dependent power-law exponent is different from the $T$-dependent power-law exponent because when the TLL and LFL exist in the same system, the response of TLL-TLL tunnelling differs from that of TLL-LFL tunnelling[27,28].

We also studied the system in the presence of a magnetic field. The scaled conductance versus scaled bias voltage curves at different temperatures coincide with each other (Fig. 2f), indicating that the Luttinger liquid exists in a 14 T magnetic field. Furthermore, from the $dI/dV$-$V_{bias}$ curves at $T$=1.6 K, $B$=14 T and different $V_{bg}$ values in Fig. 2e, the extracted exponent $\alpha$ (orange dots in Fig. 2d) also has the same dependence as a function of $V_{bg}$ regardless of the presence of a magnetic field. We repeated this phenomenon in device #2 (Supplementary Fig. 4). In particular, for device #2 which can possess dual carriers from electrons to holes crossing the CNP (Supplementary Fig. 3b), we found that $\alpha$ always increases as the system approaches the CNP regardless of which carrier, indicating that this might be due to the gate-tunable interaction strength.

**Influence from the bulk gap**

Given that Ta$_2$Pd$_3$Te$_5$ has a so complex band structure, and the magnitude of the bulk gap may also modulate the power-law exponents $α$ and $β$ of the edge Luttinger liquid state, investigating the effects of the gate voltage tuning on the bulk band gap is essential. The $T$-dependent curves at different $V_{bg}$ values coincide with each other at high temperature, indicating that the bulk gap remains unchanged despite the change in the gate voltage. We subsequently processed the $V_{bias}$-dependent curves to search for further evidence.

$V_{bias}$ can produce bias current $I_{Zener}$ by the Zener tunnelling effect[29]. The magnitude of the current $I_{Zener}$ satisfies the following:

$$I_{Zener} \sim (eV-\Delta)^{\frac{3}{2}} exp\left(-\frac{\Delta}{eV}\frac{L}{\xi}\right) \quad (1)$$

where $V$ denotes $V_{bias}$, $\Delta$ denotes the band gap, $L$ denotes the distance between the source and drain, and $\xi$ denotes the coherence length between charge carriers. A schematic of Zener tunnelling in the Ta$_2$Pd$_3$Te$_5$ system is shown in Fig. 3a; Fig. 3b shows a fitting by this formula, where the edge gap and bulk gap are extracted from the log$R$-1/$T$ plot shown in the inset. An $R^2$ value notably close to 1 indicates excellent goodness of fit, illustrating the strong applicability of the formula within our system. Even when Ta$_2$Pd$_3$Te$_5$ is reported as a possible excitonic insulator whose bulk gap continues to change at higher temperatures[19,20], the formula is still suitable[29]. The nearly twofold difference in coherence length between low-bias and high-bias regimes also indicates distinct properties of edge states versus bulk states[21].

It can be readily deduced from the $I_{Zener}$ formula that if gating has no influence on the gap size but only tunes the position of the Fermi level, the $I_{Zener}$ should exhibit only a scaling difference in magnitude, while its profile remains identical. That is, in such cases, when normalization is applied to current (or $dI/dV$) curves measured at different $V_{bg}$ values, all curves should coincide completely.

We extracted bulk-state segments from all the $dI/dV$ curves in Fig. 2c and e for normalization, as shown in Fig. 3c and d, they indeed coincide completely, which confirms that gating is not able to change the bulk gap size but can only tune the chemical potential/Fermi level. In other words, the tuning of the power-law exponents $\beta$ and $\alpha$ in the Luttinger liquid is related only to the change in edge state or, more precisely, the electron-electron interaction in these Q1D edge channels.

**Magnetic field-dependent behaviour of the Luttinger liquid**

Fig. 2d clearly shows that the magnetic field can tune the power-law exponent $\alpha$ in the edge Luttinger liquid state. To further investigate these phenomena, we measured $dI/dV$-$T$ and $dI/dV$-$V_{bias}$ curves at $V_{bg}$=10 V and a series of magnetic fields (Fig. 4a and c). The magnetic field is perpendicular to the sample plane, and a $V_{bg}$ of 10 V can be used to tune the sample to the edge Luttinger liquid state at low temperature. As shown in Fig. 4b and d, the power-law exponents $\beta$ and $\alpha$ increase as the magnetic field strength increases. This pattern is repeated near the CNP of device #1 (Supplementary Fig. 1) and at different $V_{bg}$ values of device #2 (Supplementary Fig. 5). Moreover, we measured $dI/dV$-$B$ curves for a series of $V_{bias}$ values near the CNP (Supplementary Fig. 1e), and the magnetic field has almost no influence on $dI/dV$ at high $V_{bias}$, which reaffirms that the bulk state cannot be changed by the magnetic field.

**Magnetic-field-orientation-dependent behaviour of the Luttinger Liquid**

Surprisingly, once we changed the direction of the magnetic field, the edge Luttinger liquid state could be modified. This anisotropic property is novel and incredible for a 1D system. In magnetic field direction-dependent measurements, we rotated the device in the xz-plane in a 14 T magnetic field (Fig. 5a) and measured $dI/dV$-$V_{bias}$ curves (Fig. 5b), the magnetic field is upward and $\theta$ is the angular between the magnetic field and the sample edges. The curves coincide in the bulk state regime at high $V_{bias}$ but start to divide in the edge Luttinger liquid state regime at low $V_{bias}$, indicating slight anisotropy. This could be highlighted by the differential conductance-angle ($dI/dV$-$\theta$) curves in Fig. 5c, which exhibit an anisotropy of twofold rotational symmetry at low $V_{bias}$, even an eye shape at zero $V_{bias}$, and an isotropy at high $V_{bias}$. Despite the maximum tuning amplitude being only 3%, we repeated this phenomenon near the CNP (Supplementary Fig. 2) and in device #2 (Supplementary Fig. 6), suggesting that the anisotropy is robust and intrinsic to the sample. By further investigating the $dI/dV$-$\theta$ data, we found that the differential conductance is minimized when the field is parallel to the sample surface or, more precisely, the edge but maximized when the field is perpendicular to it. This characteristic is confirmed by the magnetoresistance behaviour, which reaches its maximum in an in-plane field (Supplementary Fig. 2c).

On the other hand, the power-law exponent $\alpha$ extracted from these curves reaches a maximum in a parallel field and a minimum in a perpendicular field (Fig. 5d). It follows the pattern that a larger $\alpha$ value comes with a lower conductance, which is also shown in the effects of gating and the application of a magnetic field. Furthermore, the magnetic field orientation has little influence on the bulk state either; thus, almost 10% of the variation in the $\alpha$ value is attributed to only the change in the edge state. The plot of $\alpha$ versus $B$ and $\theta$ in Fig. 5e systematically shows the variation in the interaction of the edge Luttinger liquid state modified by the magnetic field and its direction.

**Discussion**

The power-law exponents of the Luttinger liquid are usually closely correlated with the Luttinger liquid parameter[5,30,31] $K$ by

$$\alpha \sim (K+K^{-1}-2) \quad (2)$$

and $\alpha$ decreases when $K$ approaches 1. $K$ can be described by[31]

$$K \approx \left[1+\frac{W(q=0)}{1+\pi\hbar v_F}\right]^{-1/2} \quad (3)$$

where $\hbar$ is the reduced Plank constant, $v_F$ is the Fermi velocity and $W(q=0)$ is the energy of electrostatic interactions in the long-wavelength limit. With respect to repulsive Coulombic interactions, $W(q=0)>0$ and thereby $0<K<1$, $K$ deviates from 1 as $v_F$ decreases; thus, $\alpha$ should increase.

Despite the fact that a conventional TLL system usually exhibits a gapless and linear dispersion in band structure such that its $v_F$ should be unchanged, a nonlinear band dispersion system can also carry a TLL fluid[4]. Generally, $v_F$ can be affected by the position of the Fermi level and the gap size. When gating to the CNP, $v_F$ decreases so that $\alpha$ increases (Fig. 6a). When a magnetic field is applied, the edge gap is opened by the Zeeman effect, as shown in Fig. 6c and e, which can also decrease $v_F$ and increase $\alpha$. When the magnetic field is rotating, we also find that the edge gap changes. More precisely, a parallel magnetic field will open a larger gap than a perpendicular field, as shown in Fig. 6d and f; therefore, $v_F$ decreases and $\alpha$ increases as the magnetic field direction approaches parallel to the edge (Fig. 6b).

A direction-dependent Zeeman gap can be described as an anisotropic $g$ factor, which widely exists in SOC materials[32-36]. Quantum confinement should be taken into consideration. Studies have reported that quantum confinement reduces the $g$-factor because of the quenching of OAM[37]. In quantum dots, the anisotropy of the shape can reorient the anisotropy of its OAM, thus redistributing the anisotropy of the $g$-factor[38], which also indicates that the $g$-factor in one direction can be suppressed by confinement. Therefore, in a 1D (or Q1D) system such as a nanowire, the fact that the perpendicular confinement (0D) is stronger than that parallel to the axis (1D) results in the larger *g*-factor, as well as the larger Zeeman gap in a parallel magnetic field[39]. The edge Luttinger liquid state in $Ta_2Pd_3Te_5$ is a 1D/Q1D system, thus, the quantum-confinement-dependent anisotropic *g*-factor could be the reason for its larger Zeeman gap, as well as larger *α,* in the parallel magnetic field.

**Summary**

In this work, the edge Luttinger liquid state in $Ta_2Pd_3Te_5$ was studied, and the tunable and anisotropic properties of this 1D system were determined. We found that the power-law exponents of the Luttinger liquid could be modulated by gating and the magnetic field. Furthermore, when the magnetic field is rotated, the edge Luttinger liquid exhibits anisotropy. We attribute these novel properties to the varying edge states. First, both gating and opening a Zeeman gap by a magnetic field will change the Fermi velocity of the edge state, thus affecting the power-law exponents of the Luttinger liquid. Second, anisotropic quantum confinement in a 1D system results in an anisotropic *g*-factor, leading to an anisotropic Zeeman gap and thus anisotropic power-law exponents/conductance. Our findings demonstrate that the edge TLL in $Ta_2Pd_3Te_5$ is exquisitely sensitive to both the strength and the direction of a magnetic field, providing direct experimental evidence of the interplay between electron correlations, spin-orbit coupling and quantum confinement. Because proximity-induced superconductivity at these edges, as well as the potential quantum spin hall edge states, has been observed, the ability to control the edge *α*—and thus the underlying interaction landscape—with magnetic fields, as well as the time-reversal symmetry, provides a clear pathway towards engineering magnetic-field-driven topological superconducting phases and, ultimately, Majorana-type quantum devices based on $Ta_2Pd_3Te_5$.

**Methods**

*Growth of Ta$_2$Pd$_3$Te$_5$ crystals*

Single crystals of Ta$_2$Pd$_3$Te$_5$ were synthesized by the self-flux method. Details can be found in ref.[40].

*Device fabrication*

We first coated the SiO$_2$/Si+ substrate with PMMA and patterned the back gate electrode with electron-beam lithography (EBL). The Ti/Au electrodes were deposited using a thermal evaporation instrument installed in the glove box. Afterwards, we used polydimethylsiloxane (PDMS) to exfoliate hBN/Ta$_2$Pd$_3$Te$_5$ flakes and drop them on the back gate electrode on a micropositioning stage in the glove box to construct a Ti/Au/hBN/Ta$_2$Pd$_3$Te$_5$ stack. Next, we coated the substrate with PMMA again without moving it out of the glove box. The Ti/Au source/drain electrodes were also patterned using EBL and deposited using a thermal evaporation instrument installed in the glove box. Afterwards, the metal film was lifted off, and the device was coated with PMMA once again without being moved out of the glove box.

*Electrical transport measurements*

The electrical transport measurements were performed in cryostats (a CRYOGENIC instrument with a temperature range of ~1.5-400 K). The back gate voltage $V_{bg}$ and d.c. bias were applied using a Keithley 2400 or 2612 instrument. Standard lock-in measurements were taken with a frequency of 3-30 Hz with a small a.c. excitation below the thermal excitation range without being specifically marked.

**Acknowledgements**

The work of J.S., L.L., F.Q. and G.L. was supported by the National Key Research and Development Program of China (Grant No. 2023YFA1607400 , Grant No. 2024YFA1613200), the Beijing Natural Science Foundation (Grant No. JQ23022), the National Natural Science Foundation of China (Grant Nos. 12174430 and 92365302), and the Synergetic Extreme Condition User Facility (SECUF; https://cstr.cn/31123.02.SECUF). Y.S. acknowledges support from the National Key Research and Development Program of China ( No. 2024YFA140840), the National Natural Science Foundation of China (No. U22A6005). Y.L. acknowledges support from the National Natural Science Foundation of China (Grant No. 12404154). The work of Z.D. was supported by the National Natural Science Foundation of China (Grant No. 12504561). The work of the other authors was supported by the National Key Research and Development Program of China (Grant Nos. 2019YFA0308000, 2022YFA1403800, 2023YFA1406500, and 2024YFA1408400), the National Natural Science Foundation of China (Grant Nos. 12274436, and 12274459), the Beijing Natural Science Foundation (Grant No. Z200005), and the Synergetic Extreme Condition User Facility (SECUF, https://cstr.cn/31123.02.SECUF). The work is also funded by the Chinese Academy of Sciences President's International Fellowship Initiative (Grant No. 2024PG0003).

**Author contributions**

J.S. conceived and designed the experiments. X.G., A.W., Y.L., G.L., Z.Z., X.S. and X.D. fabricated the devices and performed the transport measurements, which were supervised by Z. D, Z.W., G.L., F. Q, T.Q., L.L., and J.S.; X.D. and Y.S. grew bulk $Ta_2Pd_3Te_5$ crystals. X.G. and J.S. analysed the data and wrote the manuscript, with input from all the authors.

**Competing interests**

The authors declare that they have no competing interests.

**Data availability**

All the data needed to evaluate the conclusions in the paper are present in the main text and/or the Supplementary Information. The authors declare that all the raw data generated in this study have been deposited in Figshare (We will attach the address after the article is received).

# Figures

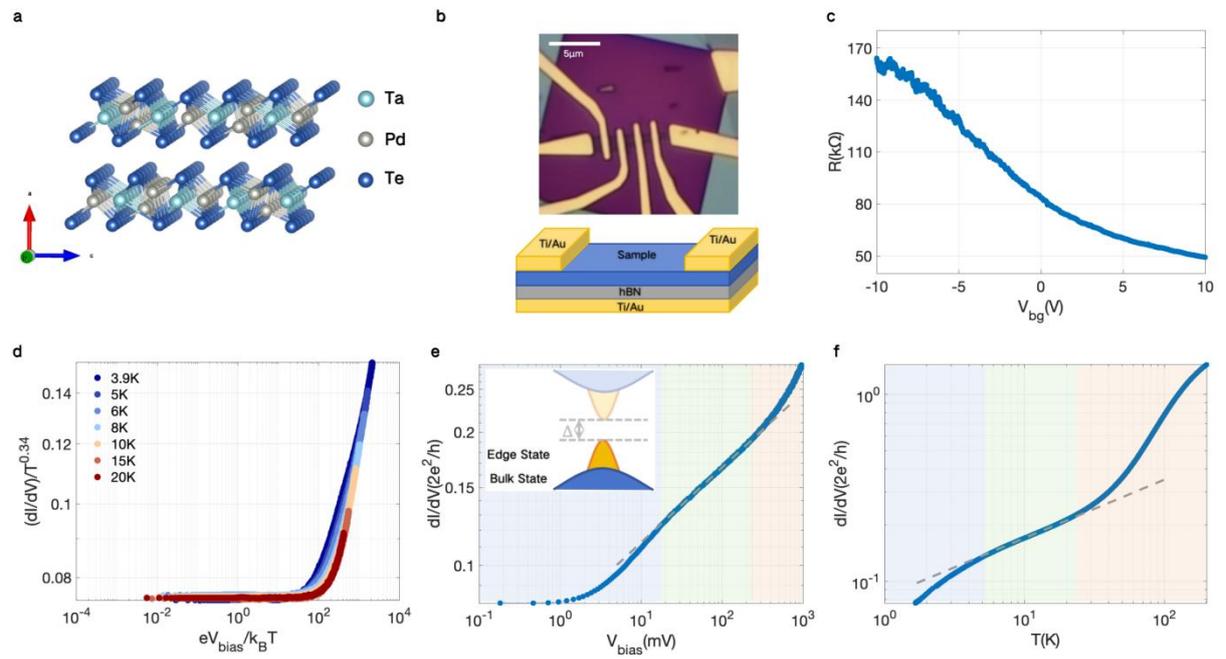

**Fig. 1. Properties of Ta$_2$Pd$_3$Te$_5$. a.** Crystal structure of Ta$_2$Pd$_3$Te$_5$. **b.** Optical photo and sketch of device #1. Scale bar, 5 μm. **c.** Differential resistance *dV/dI* versus back gate voltage $V_{bg}$ at *T*=1.6 K. **d.** Scaled conductance *(dI/dV)/T$^{0.34}$* versus scaled bias voltage $eV_{bias}/k_BT$ at $V_{bg}$=-10 V and *T*=3.9~20 K. **e.** Log-log plot of *dI/dV* versus $V_{bias}$ at $V_{bg}$=-10 V and *T*=1.6 K; **inset**: schematic of the band structure of Ta$_2$Pd$_3$Te$_5$. **f.** Log-log plot of *dI/dV* versus *T* at $V_{bg}$=-10 V. In **e** and **f**, the blue/green/orange regions of the background show the edge gap/edge Luttinger liquid/bulk semiconductor behaviour, and the grey dashed lines are linear fits of the Luttinger liquid segment in the log-log plot to show its power-law behaviour.

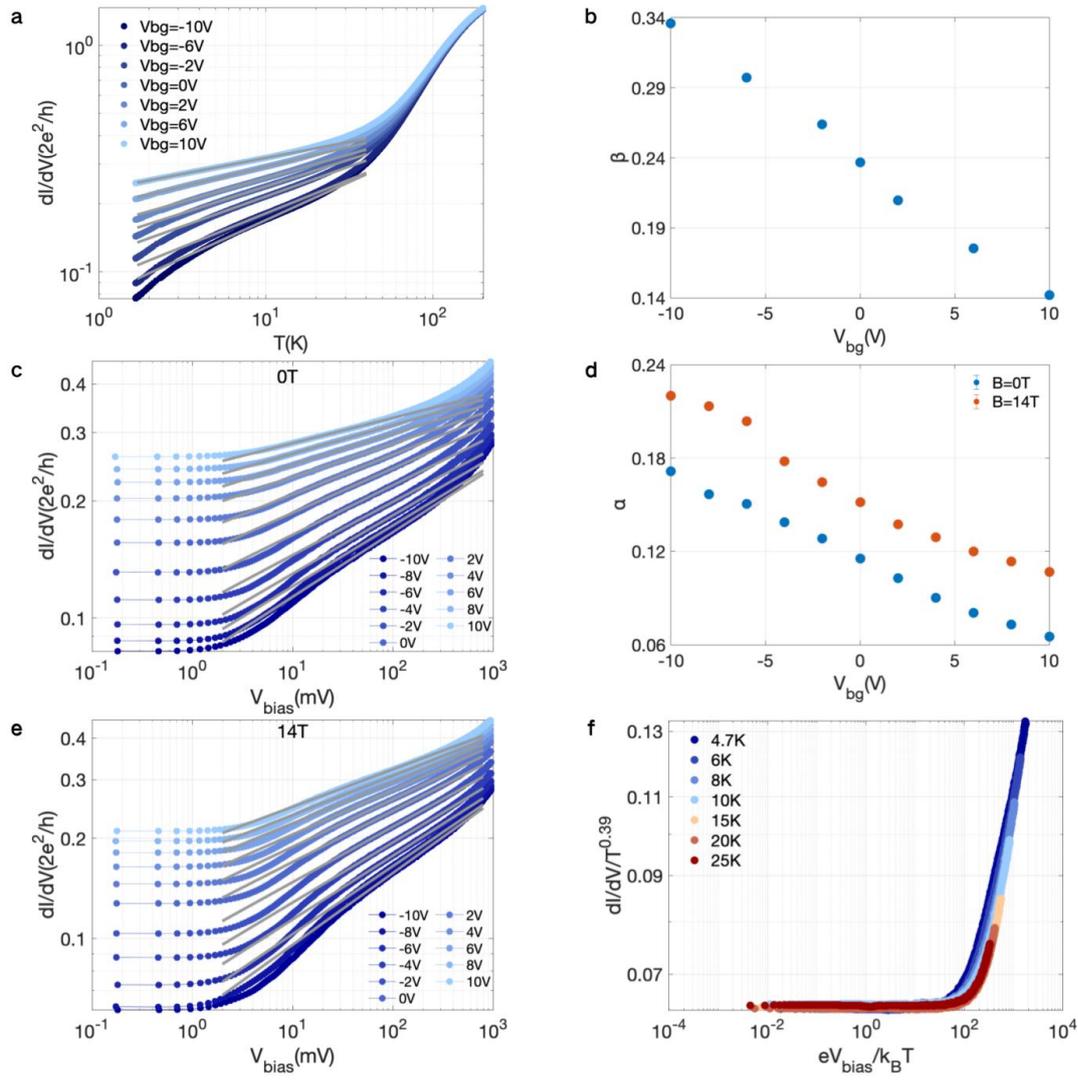

**Fig. 2. Gate Tunability of the Luttinger liquid. a.** Log-log plot of *dI/dV* versus *T* for a series of $V_{bg}$. **c and e.** Log-log plot of *dI/dV* versus $V_{bias}$ at *T*=1.6 K and a series of $V_{bg}$ when *B*=0 T/14 T. In **a, c and e.** The grey lines are linear fits of the log-log scaled power-law behaviour in the edge Luttinger liquid state. **b.** Power-law exponents *β* extracted from **a**. **d.** Power-law exponents *α* extracted from **c and e**. The blue and orange dots are *α* at *B*=0 T and 14 T, respectively. **f.** Scaled conductance *(dI/dV)/$T^{0.39}$* versus scaled bias voltage $eV_{bias}/k_BT$ measured near the CNP at $V_{bg}$=-10 V, B=14 T and T=4.7~25 K. In **b and d.** The error bars are determined by the 95% confidence interval of the fitting coefficients and are too short and covered by dots, the same as the following figures.

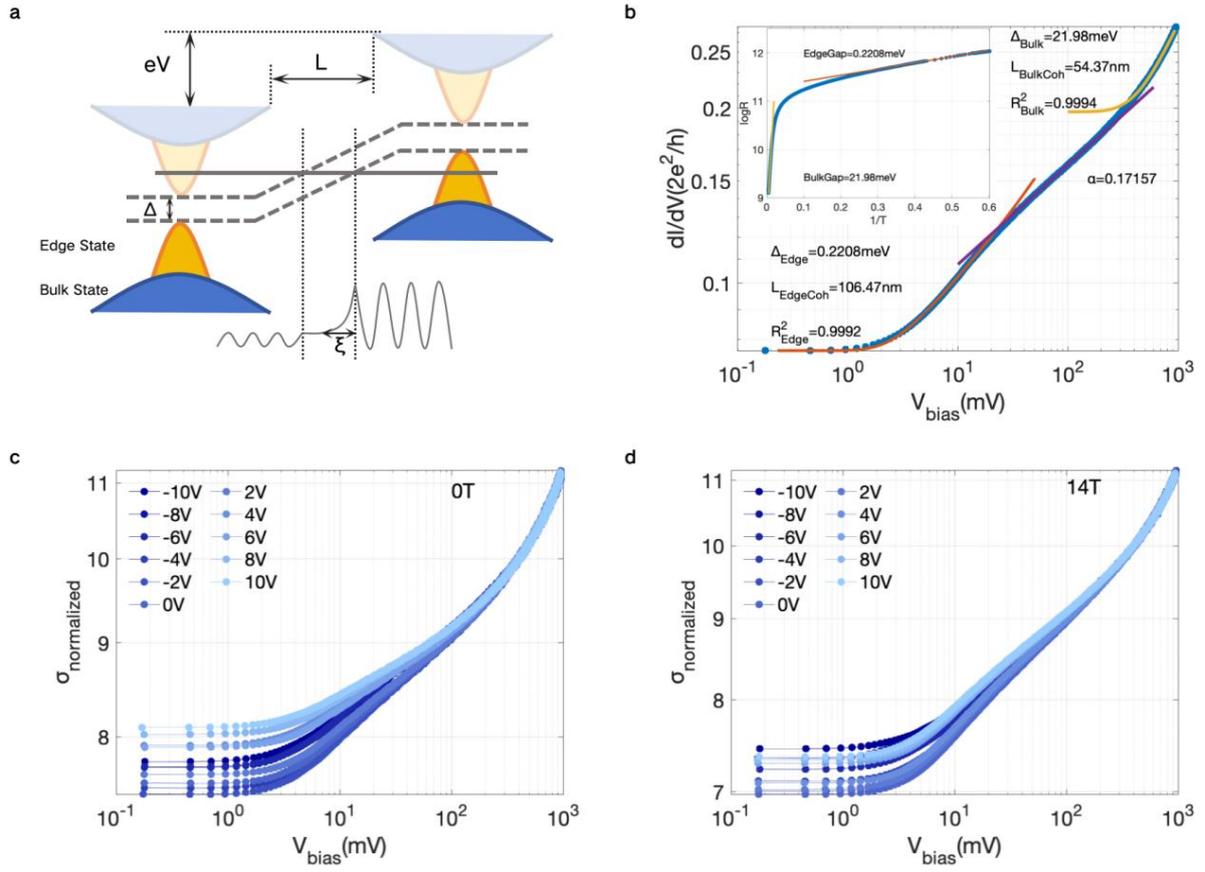

**Fig. 3. Zener tunnelling and normalization of current curves. a.** Schematic of Zener tunnelling in the Ta$_2$Pd$_3$Te$_5$ system. **b.** Zener tunnelling fitting of $dI/dV$ -$V_{bias}$ curves at $V_{bg}$=-10 V, $B$=0 T and $T$=1.6 K; the blue dots are the measured data, the orange/yellow lines are the Zener tunnelling fitting of the edge gap/bulk semiconductor regime, respectively, and the purple line is the linear fitting of the log-log scaled power-law behaviour in the edge Luttinger liquid state. **Inset**: log$R$-1/$T$ plot at $V_{bg}$=-10 V and $B$=0 T; the orange and yellow lines represent edge gap and bulk gap fitting, respectively; the edge gap and the bulk gap are extracted here by $\Delta = k \cdot 2k_B$, where $k$ is the slope of the low and high temperature region of the log$R$-1/$T$ curve. **c and d.** Normalization of $dI/dV$ curves in Fig. **2c and e**. We chose $V_{bias}$=400 mV and 900 mV as reference point, $\sigma_{normalized}=\frac{\sigma - \sigma_{400mV}}{\sigma_{900mV} - \sigma_{400mV}}+10$.

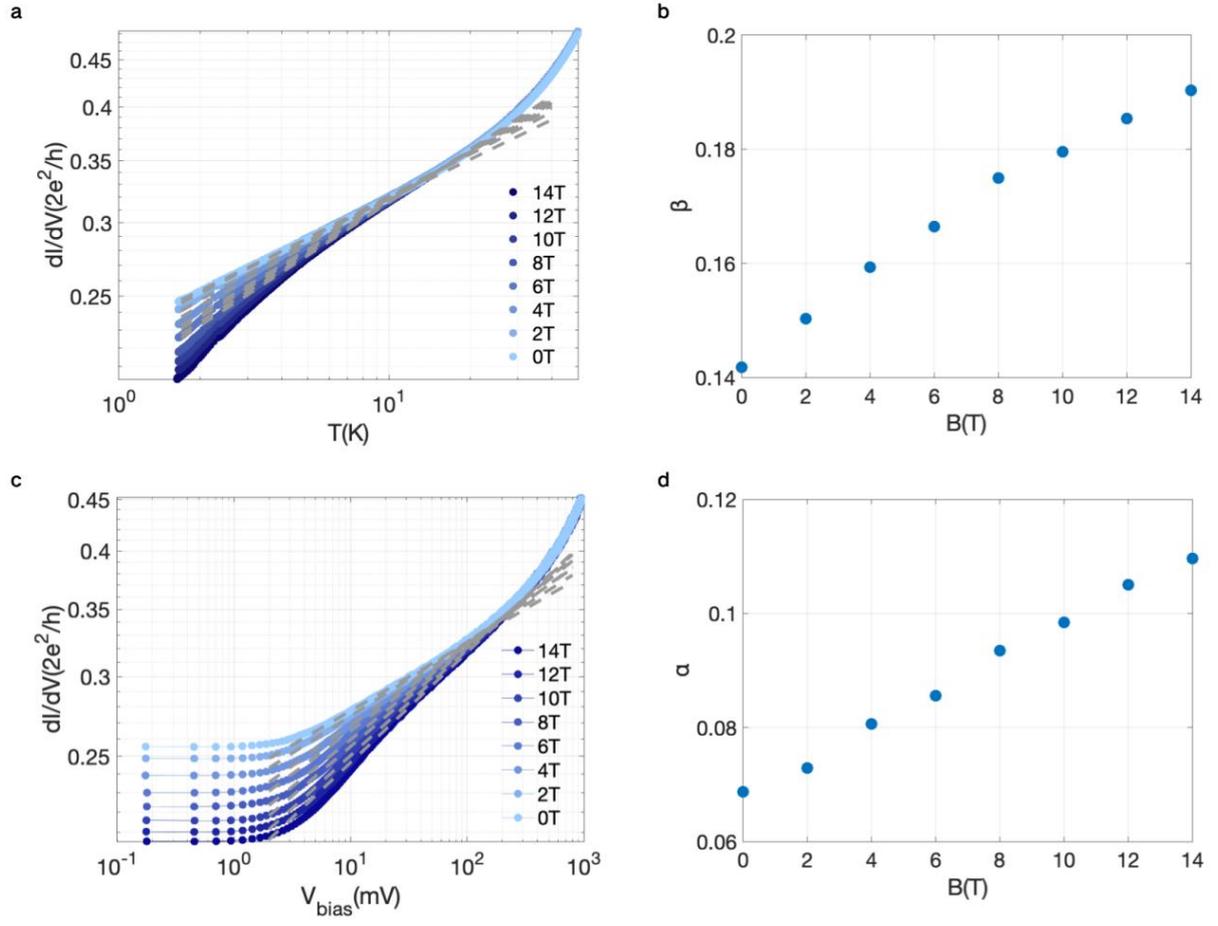

**Fig. 4. Magnetic-field dependence of the Luttinger liquid. a.** Log-log plot of *dI/dV* versus *T* at $V_{bg}$=10 V and different *B*. **c.** Log-log plot of *dI/dV* versus $V_{bias}$ at $V_{bg}$=10 V, *T*=1.6 K and different *B*. In **a and c.** The grey dashed lines are linear fits of the log-log scaled power-law behaviour in the edge Luttinger liquid state. **b and d.** Power-law exponents *β* and *α* extracted from **a and c**.

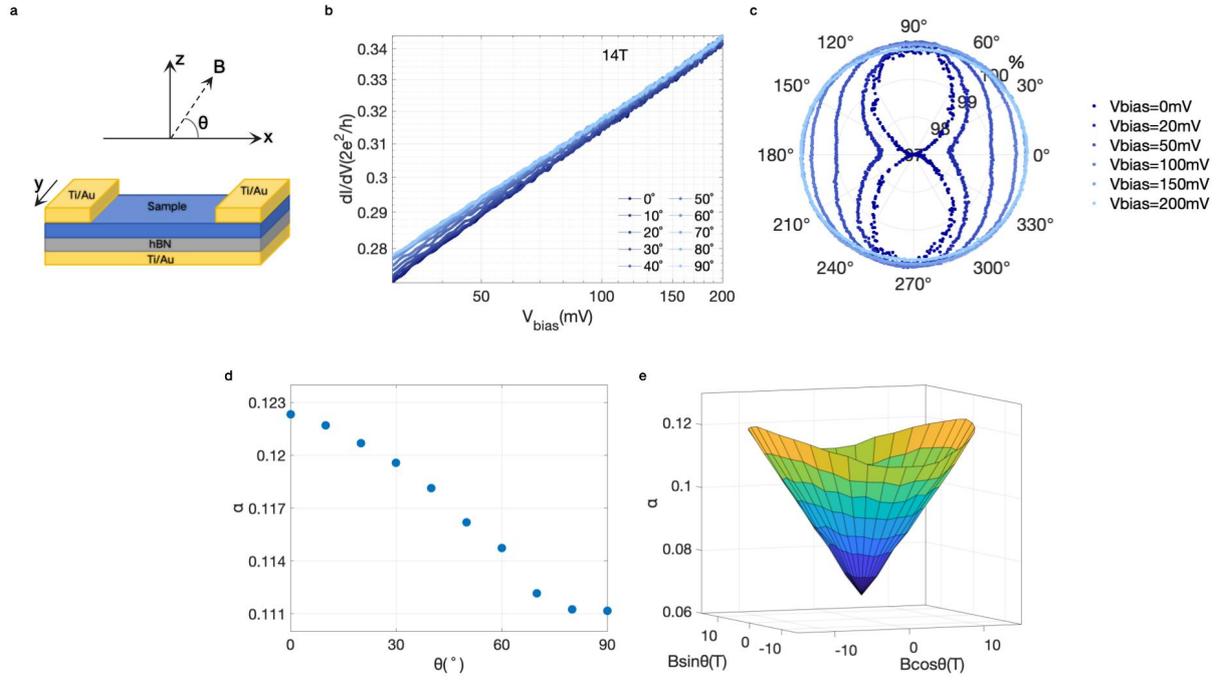

**Fig. 5. Dependence of the Magnetic Field Orientation of a Luttinger Liquid. a.** Schematic of the magnetic field orientation. **b.** Log-log plot of *dI/dV* versus $V_{bias}$ in different directions of the magnetic field at *B*=14 T, $V_{bg}$=10 V and *T*=1.6 K. **c.** Tuning amplitude of *dI/dV* versus *θ* at *B*=14 T, $V_{bg}$=10 V and *T*=1.6 K. $amplitude=\frac{dI/dV}{(dI/dV)_{max}}\times 100\%$. **d.** Power-law exponent *α* extracted from the edge Luttinger liquid regime of **b**. **e.** *α* versus *B* & *θ* at $V_{bg}$=10 V and *T*=1.6 K. Image in the region 90º< *θ*<360º is derived from the image in the region 0º≤ *θ*≤90º through symmetry.

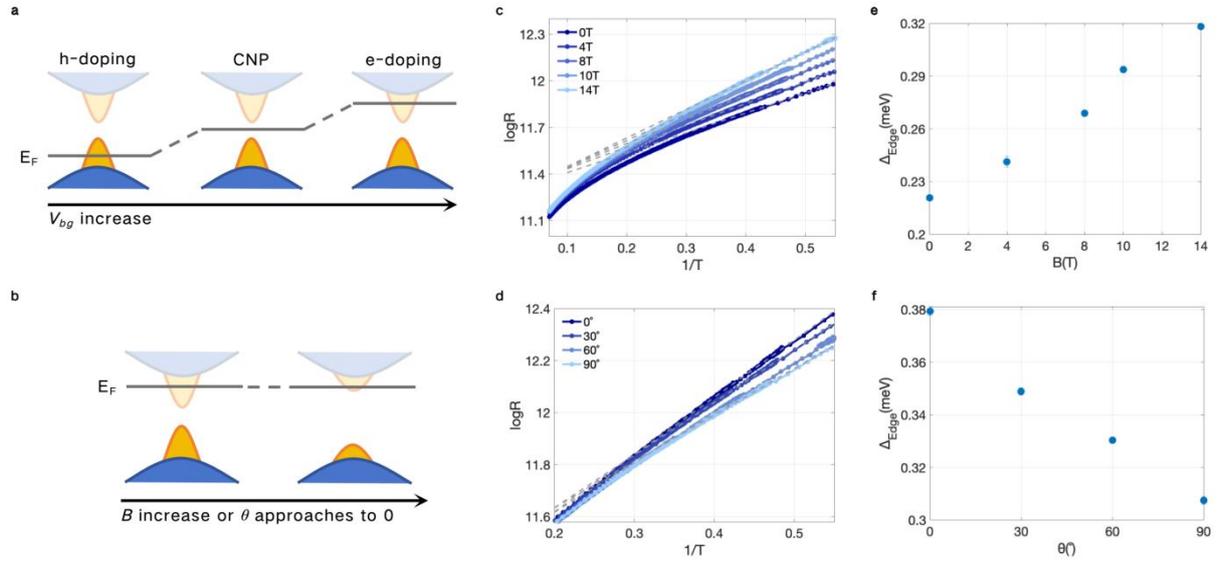

**Fig. 6. Mechanism of the tunable edge Luttinger liquid. a.** Schematic of gate tuning. $v_F$ decreases as the system approaches the CNP. **b.** Schematic of magnetic and direction tuning. If the magnetic field increases or approaches parallel to the edge, a larger Zeeman gap can be opened, thus decreasing $v_F$. **c.** log$R$-1/$T$ plot at the CNP to fit the edge gap for different perpendicular magnetic fields. **d.** log$R$-1/$T$ plot at $B$=12T and CNP to fit the edge gap in different directions. **e and f.** Edge gap extracted from the low temperature region of **c and d** by $\Delta = k \cdot 2k_B$, where $k$ is the slope of the log$R$-1/$T$ curve.

**Supplementary information**

**Magnetic-Field Control of Tomonaga-Luttinger Liquids in Ta$_2$Pd$_3$Te$_5$ Edge States**

Xingchen Guo[1,2,†], Anqi Wang[1,2,†], Xiutong Deng[1,2,†], Yupeng Li[3], Guo'an Li[1,2], Zhiyuan Zhang[1,2], Xiaofan Shi[1,2], Xiao Deng[1,2], Ziwei Dou[1], Guangtong Liu[1], Fanming Qu[1,2], Zhijun Wang[1,2], Tian Qian[1], Youguo Shi[1,2,*], Li Lu[1,2,*], Jie Shen[1,*]

[1]Beijing National Laboratory for Condensed Matter Physics, Institute of Physics, Chinese Academy of Sciences, Beijing 100190, China

[2]School of Physical Sciences, University of Chinese Academy of Sciences, 100049 Beijing, China

[3]Hangzhou Key Laboratory of Quantum Matter, School of Physics, Hangzhou Normal University, Hangzhou 311121, China

†These authors contributed equally to this work

**\*Corresponding author**

e-mail: ygshi@iphy.ac.cn; lilu@iphy.ac.cn; shenjie@iphy.ac.cn

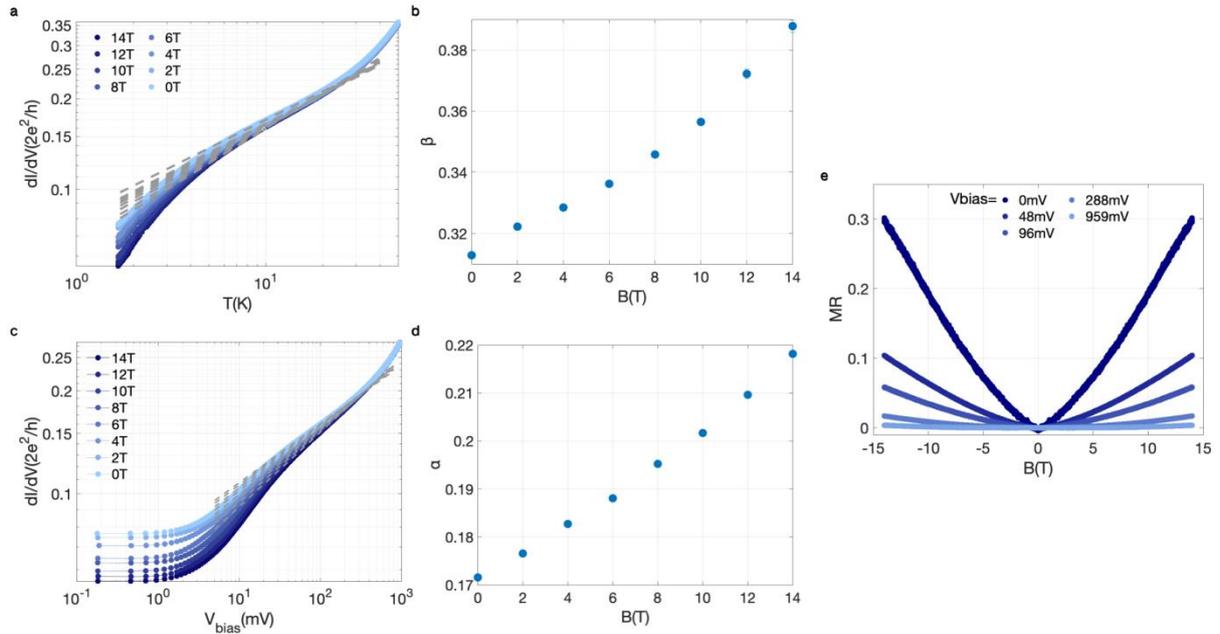

**Supplementary Fig. 1. Magnetic-field dependence of Luttinger liquid at CNP in Device #1. a.** Log-log plot of *dI/dV* versus *T* at $V_{bg}$=-10 V and different magnetic fields *B*. **c.** Log-log plot of *dI/dV* versus $V_{bias}$ at $V_{bg}$=-10 V, *T*=1.6 K and different *B*. In **a and c.** Grey dashed lines are linear fits of the log-log scaled power-law behaviour in the edge Luttinger liquid state. **b and d.** Power-law exponents *α* and *β* extracted from **a and c**. **e.** MR curves at $V_{bg}$=-10 V and *T*=2 K.

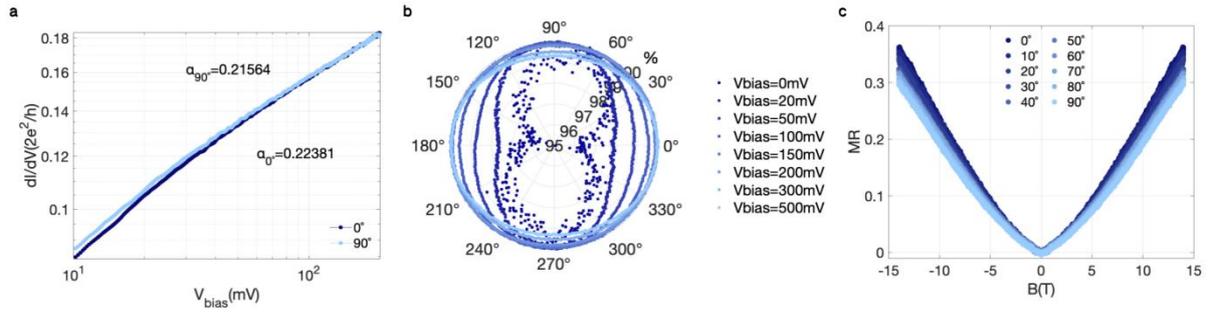

**Supplementary Fig. 2. Magnetic Field Orientation Dependence of Luttinger Liquid at CNP in Device #1 and bulk. a.** Log-log plot of *dI/dV* versus $V_{bias}$ at $V_{bg}$=-10 V, B=14 T, *T*=1.6 K and different directions. **b.** Tuning amplitude of *dI/dV* versus $\theta$ at $V_{bg}$=-10 V and different $V_{bias}$. $amplitude = \frac{dI/dV}{(dI/dV)_{max}} \times 100\%$. **c.** *MR* versus *B* at $V_{bg}$=-10 V, *T*=2 K and different directions of Device #1. $MR = \frac{R_{xx}}{R_{xx}(B=0)} - 1$.

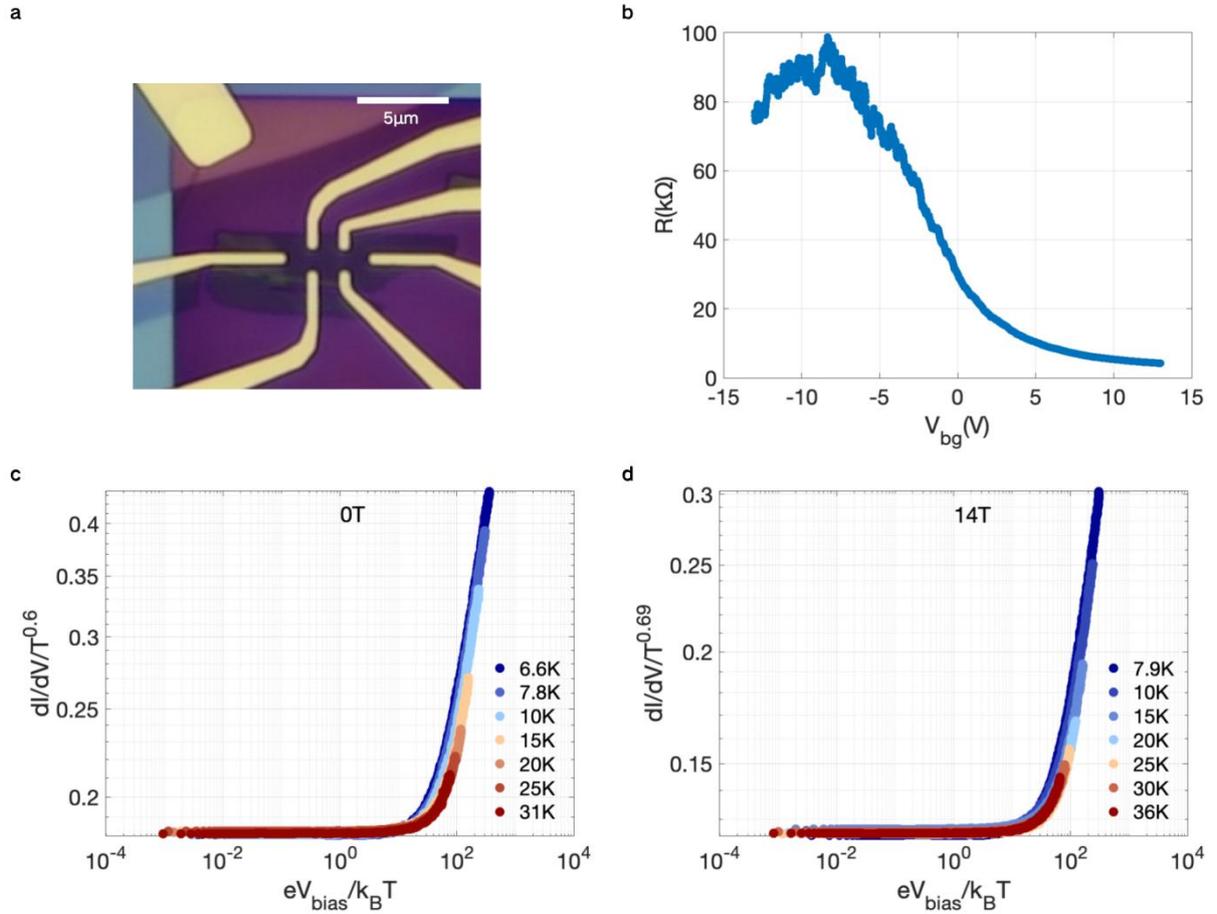

**Supplementary Fig. 3. Basic properties of Device #2. a.** A photo of device #2. Scale bar, 5 μm. **b.** Differential resistance $dV/dI$ versus back gate voltage $V_{bg}$ measured at $T$=1.6 K. This indicates that the CNP is near $V_{bg}$=-8 V. **c and d.** Scaled conductance $(dI/dV)/T^\beta$ versus scaled bias voltage $eV_{bias}/k_BT$ at CNP and $B$=0 T/14 T, respectively; $\beta$=0.6 when $B$=0 T; $\beta$=0.69 when $B$=14 T. Lines measured at different temperatures coincide with each other; this phenomenon indicates that despite the presence of a magnetic field, there is a Luttinger liquid in device #2.

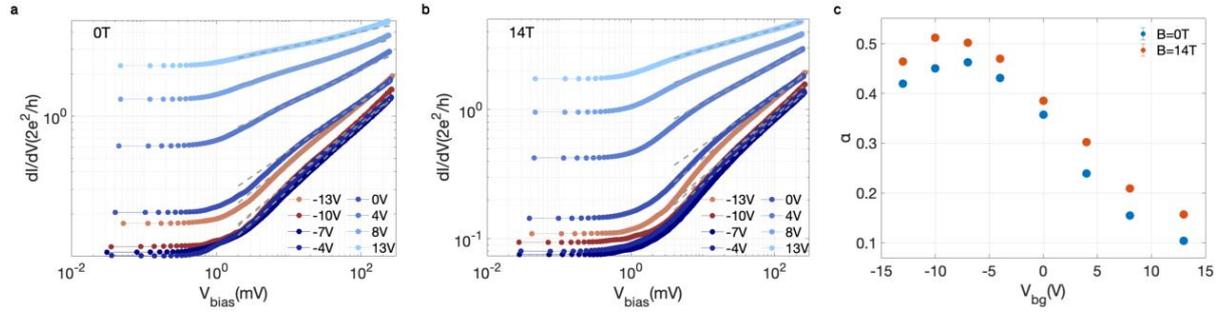

**Supplementary Fig. 4. Gate Tunability of the Luttinger liquid in Device #2. a and b.** Log-log plot of *dI/dV* versus $V_{bias}$ at *T*=1.6 K and *B*=0 T/14 T; the blue lines are measured in the electron-doping regime, the red lines are measured in the hole-doping regime, and the grey dashed lines are linear fits of the log-log scaled power-law behaviour in the edge Luttinger liquid state. **c.** Power-law exponent *α* extracted from **a and b**.

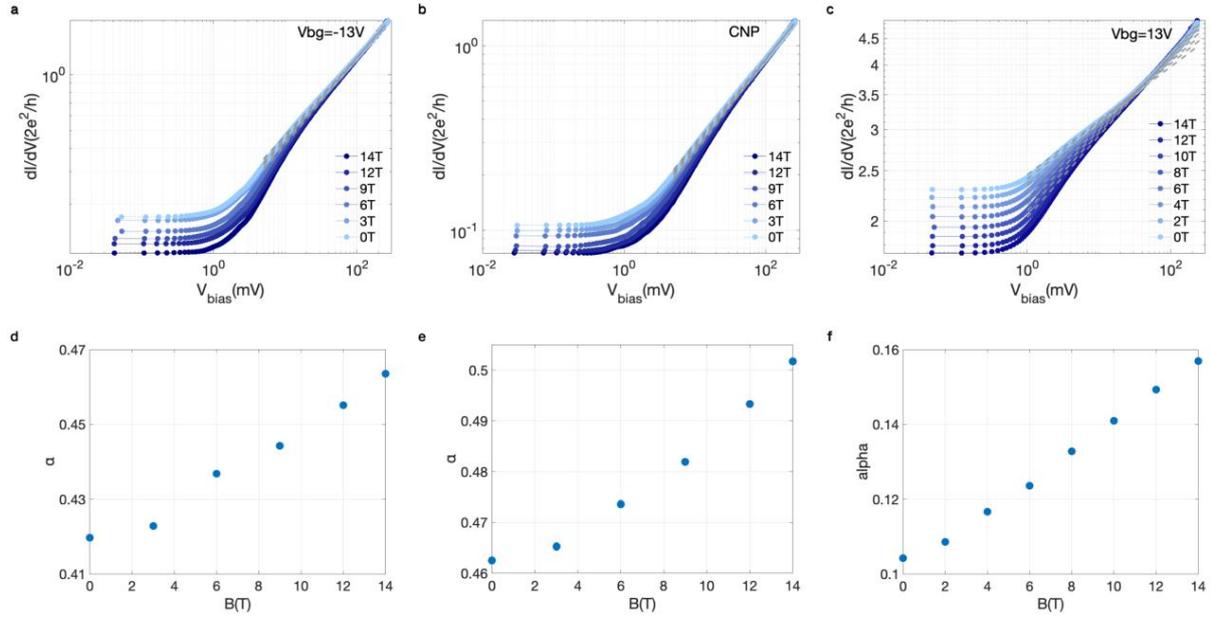

**Supplementary Fig. 5. Magnetic-field dependence of Luttinger liquid in Device#2. a~c.** Log-log plot of *dI/dV* versus $V_{bias}$ at *T*=1.6 K and $V_{bg}$=-13 V, -8 V, and 13 V, respectively, where $V_{bg}$=-13 V is in hole-doping regime, $V_{bg}$=-8 V is at CNP and $V_{bg}$=13 V is in electron-doping regime. Grey dashed lines are linear fits of the log-log scaled power-law behaviour in the edge Luttinger liquid state. **d~f** Power-law exponent *α* extracted from **a~c**.

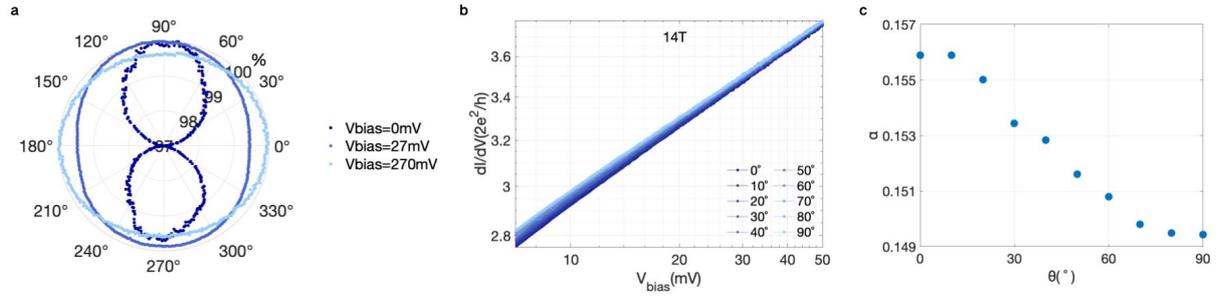

**Supplementary Fig. 6. Magnetic Field Orientation Dependence of Luttinger Liquid in Device #2. a.** Tuning amplitude of *dI/dV* versus $\theta$ at $V_{bg}$=13 V and *B*=14 T. $amplitude=\frac{dI/dV}{(dI/dV)_{max}}\times 100\%$. **b.** Log-log plot of *dI/dV* versus $V_{bias}$ in different direction at *T*=1.6 K, $V_{bg}$=13 V, and *B*=14 T. **c.** Power-law exponent *α* extracted from the edge Luttinger liquid region of **b**.